\documentclass{eptcs}
\providecommand{\event}{LFMTP 2020} 
\usepackage{breakurl}             
\usepackage{underscore}           

\usepackage{xcolor}

\usepackage{proof}
\usepackage{stmaryrd}
\usepackage[all]{xy}
\usepackage{amssymb}
\usepackage{amsmath}
\usepackage{amsthm}
\usepackage[inline]{enumitem}
\usepackage[normalem]{ulem}
\usepackage{wasysym}
\usepackage{multicol}

\theoremstyle{definition}

\newtheorem*{example}{Example}

\newcommand{\Set}{\mathbf{Set}}

\newcommand{\Nat}{\mathbb{N}}

\newcommand{\dsem}[1]{\{\!\!\!\{ #1 \}\!\!\!\}}

\newcommand{\I}{\mathsf{I}}
\newcommand{\ot}{\otimes}
\newcommand{\lo}{\multimap}

\newcommand{\loL}{\mathbin{{\multimap}{\mathsf{L}}}}
\newcommand{\loC}{\mathbin{{\multimap}{\mathsf{C}}}}
\newcommand{\loR}{\mathbin{{\multimap}{\mathsf{R}}}}
\newcommand{\loe}{\mathbin{{\multimap}}{\mathsf{e}}}
\newcommand{\loi}{\mathbin{{\multimap}}{\mathsf{i}}}

\newcommand{\SCxt}{\mathbf{SCxt}}
\newcommand{\Cxt}{\mathbf{Cxt}}

\newcommand{\C}{\mathbb{C}}
\newcommand{\Cop}{\C^{\mathsf{op}}}
\newcommand{\D}{\mathbb{D}}

\newcommand{\inv}{\textsf{I}}

\newcommand{\n}{{-}}
\newcommand{\Var}{\mathsf{At}}
\newcommand{\At}{\mathsf{At}}

\newcommand{\tto}{\Longrightarrow}
\newcommand{\Tm}{\mathsf{Fma}}
\newcommand{\Fma}\Tm
\newcommand{\id}{\mathsf{id}}
\newcommand{\comp}{\circ}

\newcommand{\plug}{\circ_0}

\newcommand{\ax}{\mathsf{ax}}
\newcommand{\uf}{\mathsf{pass}}
\newcommand{\pass}{\uf}

\newcommand{\ufinv}{\mathsf{act}}

\newcommand{\scut}{\mathsf{scut}}
\newcommand{\ccut}{\mathsf{ccut}}
\newcommand{\ccutj}{\mathsf{ccut}_{\mathsf{Fma}}}

\newcommand{\sound}{\mathsf{sound}}
\newcommand{\cmplt}{\mathsf{cmplt}}

\newcommand{\focus}{\mathsf{focus}}
\newcommand{\emb}{\mathsf{emb}}
\newcommand{\hered}{\mathsf{hered}}
\newcommand{\eval}{\mathsf{eval}}
\newcommand{\reify}{\mathsf{reify}}
\newcommand{\reflect}{\mathsf{reflect}}
\newcommand{\nbe}{\mathsf{nbe}}
\newcommand{\netoF}{\mathsf{ne2F}}
\newcommand{\Ftone}{\mathsf{F2ne}}
\newcommand{\nftoI}{\mathsf{nf2I}}
\newcommand{\Itonf}{\mathsf{I2nf}}

\newcommand{\Lstar}{L^{\star}}

\renewcommand{\doteqdot}{\mathrel{\circeq_\mathsf{nd}}}

\newcommand{\proofbox}[1]{\begin{tabular}{c} #1 \end{tabular}}

\newcommand{\asem}[2]{\llbracket #1 | #2 \rrbracket}

\newcommand{\FskPCl}{\mathbf{FSkPCl}}
\newcommand{\FSkPCl}{\FskPCl}

\newcommand{\stseq}[3]{#1 \mid #2 \longrightarrow #3}
\newcommand{\stseqname}[4]{#1 \mid #2 \stackrel{#3}{\longrightarrow} #4}

\newcommand{\stseqnd}[3]{#1 \mid #2 \longrightarrow_\mathsf{nd} #3}

\newcommand{\G}{\mathsf{DC}}
\newcommand{\M}{\mathbb{M}}

\newcommand{\jY}{\widehat{j}}
\newcommand{\iY}{\widehat{\imath}}
\newcommand{\LY}{\widehat{L}}

\newcommand\nf[3]{#1 \mid #2 \longrightarrow_{\mathsf{nf}} #3}
\renewcommand\ne[3]{#1 \mid #2 \longrightarrow_{\mathsf{ne}} #3}
\newcommand\ndp[3]{#1 \mid #2 \longrightarrow_{\mathsf{p}} #3}

\newcommand{\dcomp}{\mathsf{comp}}
\newcommand\stseqI[4][]{#2 \mid #3 \overset{#1}{\longrightarrow_{\mathsf{I}}} #4}
\newcommand\stseqF[4][]{#2 \mid #3 \overset{#1}{\longrightarrow_{\mathsf{F}}} #4}
\newcommand\stseqP[4][]{#2 \mid #3 \overset{#1}{\longrightarrow_{\mathsf{P}}} #4}
\newcommand\stseqk[4][]{#2 \mid #3 \overset{#1}{\longrightarrow_k} #4}

\newcommand\stseqM[4][]{\M(#2 | #3; #4)}
\newcommand\stseqG[4][]{\G(#2 | #3; #4)}

\newcommand\stseqMpr[4][]{\M'(#2 | #3; #4)}
\newcommand\stseqU[4][]{(U\, \C)(#2 | #3; #4)}

\newcommand{\loosen}{\mathsf{loosen}}

\newcommand{\cxt}{\mathsf{cxt}}
\newcommand{\scxt}{\mathsf{scxt}}

\newcommand{\tu}[1]{}
\newcommand{\nv}[1]{}
\newcommand{\nz}[1]{}

\title{Deductive Systems and Coherence for \\ Skew Prounital Closed Categories}
\author{Tarmo Uustalu
\institute{Reykjavik University, Reykjavik, Iceland}
\institute{Tallinn University of Technology, Tallinn, Estonia}
\email{tarmo@ru.is}
\and
Niccol{\`o} Veltri
\institute{Tallinn University of Technology, Tallinn, Estonia}
\email{niccolo@cs.ioc.ee}
\and
Noam Zeilberger
\institute{{\'E}cole Polytechnique, Palaiseau, France}
\email{noam.zeilberger@lix.polytechnique.fr}
}
\def\titlerunning{Deductive Systems and Coherence for Skew Prounital Closed Categories}
\def\authorrunning{T. Uustalu, N. Veltri \& N. Zeilberger}
\begin{document}
\maketitle

\begin{abstract}

  In this paper we develop the proof theory of skew prounital closed
  categories. These are variants of the skew closed
  categories of Street where the unit is not represented. Skew closed 
  categories in turn are a weakening of the closed
  categories of Eilenberg and Kelly where no structural law is
  required to be invertible. The presence of a monoidal structure in
  these categories is not required.  We construct several equivalent
  presentations of the free skew prounital closed category on a given
  set of generating objects: a categorical calculus (Hilbert-style system),
  a cut-free sequent calculus and a natural deduction system corresponding to
a variant of 
  planar (= non-commutative linear) typed lambda-calculus. We solve the
  coherence problem for skew prounital closed categories by showing
  that the sequent calculus admits focusing and presenting two
  reduction-free normalization procedures for the natural deduction calculus:
  normalization by evaluation and hereditary substitutions. Normal natural deduction derivations ($\beta\eta$-long forms) are in one-to-one
  correspondence with derivations in the focused sequent
  calculus. Unexpectedly, the free skew prounital closed category on a
  set satisfies a left-normality condition which makes it lose its
  skew aspect. This pitfall can be avoided by considering the free
  skew prounital closed category on a skew multicategory instead. The
  latter has a presentation as a cut-free sequent calculus
  for which it is easy to see that the left-normality
  condition generally fails.

  The whole development has been fully formalized in the
  dependently typed programming language Agda.
\end{abstract}

\section{Introduction}
\label{sec:intro}

Proof theory and category theory have gone hand in hand since the
pioneering works of Lambek~\cite{Lam:ded1,Lam:ded2,Lam:ded3} and a number of
researchers that followed immediately, like Lawvere~\cite{Law}, 
Szabo~\cite{Sza:catep,Sza:algp}, Mann \cite{Man:conepc},
Mints~\cite{Min:cloctp}. Category theory helps the proof theorist with
mathematical models for logical proof systems, which should help
tackling problems like analysis of the connections between different
types of proof systems, e.g., sequent calculi and natural deduction
\cite{Zuc:corcen}. On the other hand, proof theory provides the
category theorist with a toolbox for identifying the internal language
of categories and for solving problems of combinatorial nature such as
Mac Lane's coherence problem~\cite{ML:natac,Kel:maclcc}.

Given a certain notion of category with structure, it is natural to
ask whether there exists deductive systems (with good proof-theoretic
properties) presenting the ``canonical'' category with that
structure. For Cartesian closed categories, such systems are given by,
e.g., the sequent calculus of intuitionistic logic and its natural
deduction system, which we also know as typed lambda-calculus. For
symmetric monoidal closed categories, some such systems are the
sequent calculus of intuitionistic linear logic (with
$\I,{\ot},{\lo}$) and the linear variant of typed lambda-calculus
\cite{BBHdP,Tro:natdll}. Formally, the ``canonical'' category with
structure arises from a \emph{free} construction. E.g., simply-typed
lambda-calculus (with $1,{\times},{\Rightarrow}$) and with atomic
types taken from a set $\At$, is a presentation of the free Cartesian
monoidal closed category on $\At$.

In recent work, we have investigated the deductive systems associated
to \emph{skew monoidal categories}~\cite{UVZ:seqcsm,UVZ:protpn}. These
are a weakening, first studied by Szlach{\'a}nyi~\cite{Szl:skemcb}, of
\emph{monoidal categories}~\cite{Ben:catam,ML:natac} in which the
unitors and associator are not required to be invertible, they are
merely natural transformations in a particular direction. These
categories are not uncommon, e.g., they appear in the study of relative
monads~\cite{ACU:monnnb} and quantum categories~\cite{LS:skemsw}.  The
free skew monoidal category on a set $\At$ can be constructed as a
sequent calculus with sequents of the form $\stseq S{\Gamma} C$, where
the antecedent is split into an optional formula $S$, called the
\emph{stoup}, and a list of formulae $\Gamma$, the
\emph{context}. This sequent calculus has some peculiarities: left
rules apply only to the formula in the special stoup position, while
the tensor right rule forces the formula in the stoup of the
conclusion to be the formula in the stoup of the first premise. This
sequent calculus enjoys cut elimination and a focused subsystem,
defining a root-first proof search strategy attempting to build a
derivation of a sequent. The focused calculus finds exactly one
representative of each equivalence class of derivations and is thus a
concrete presentation of the free skew monoidal category, as such
solving the coherence problem for skew monoidal categories.

In this paper, we perform a similar proof-theoretic analysis of
\emph{skew prounital closed categories}~\cite{UVZ:eilkr}. These are
the skew variant of \emph{prounital closed category} of
Shulman~\cite[Rev.~49]{ncatlab:cloc}, which in turn is a relaxation of
the notion of \emph{closed category} by Eilenberg and
Kelly~\cite{EK:cloc,Lap:cohncc}. Intuitively, a prounital closed
category is a category with an internal hom object $A \lo B$
for any two objects $A$ and $B$. There is no requirement
for a unit object $\I$, nor for a tensor product $\ot$. But the unit
is implicitly present to a degree thanks to the presence of a functor
$J : \C \to \Set$, with $J\, A$ playing the role of the set of maps
from the non-represented unit to the object $A$. In other words, for a
closed category $\C$, we have $J \,A = \C(\I, A)$. Related categories
where the unit is ``half-there'' appear in the study of categorical
models of classical linear logic~\cite{Hou:linlwu}. Yet weaker are
Lambek's \emph{residuated categories}~\cite{Lam:ded1} (with one
implication) where the unit is completely absent. \emph{Skew closed
  categories}, a variant of closed categories where no structural law
is required to be invertible, were first considered by
Street~\cite{Str:skecc}.


We present several deductive systems giving different but equivalent
presentations of the free skew prounital closed category on a set
$\At$: a categorical calculus (Hilbert-style system), 
a cut-free sequent calculus and a natural deduction
calculus. Similarly to the skew monoidal case~\cite{UVZ:seqcsm},
sequents in sequent calculus have the form $\stseq S{\Gamma} C$ and
the left implication rule only applies to the formula in the stoup
position. The natural deduction calculus is, under Curry-Howard
correspondence, a variant of planar typed
lambda-calculus~\cite{Abr:temper,Zei:theltf}. Lambda-terms in this calculus have
all free and bound variables used exactly once and in the order of
their declaration.

We give two equivalent calculi of normal forms: a focused sequent
calculus and normal natural deduction derivations, corresponding to
canonical representatives of $\beta\eta$-equality. We show three
reduction-free (in the sense that we do not use techniques from
rewriting theory) normalization procedures:
focusing~\cite{And:logpfp}, sending a sequent calculus derivation to a
focused derivation; normalization by hereditary
substitutions~\cite{WCPW:conlfp,KA:hersst}, sending a natural
deduction derivation to a focused derivation;
normalization by evaluation~\cite{BS:inveft,AHS:catrrf}, sending a
natural deduction derivation to a normal natural deduction
derivation.

Using our sequent calculus, it is possible to show that the free skew
prounital closed category on a set satisfies a \emph{left-normality}
condition. The structural law $\jY$, that we are not asking to be
invertible, turns out to be invertible anyway. This degeneracy implies
that our sequent calculus and natural deduction calculus admit a
stoup-free presentation. In particular, the natural deduction calculus
is equivalent, under Curry-Howard correspondence, to (non-skew) planar
typed lambda-calculus. From a category-theoretic point of view, there
is no reason to construct the free skew prounital closed category on a
\emph{set} instead of a more interesting category. We conclude this
paper by discussing two equivalent presentations of the free skew
prounital closed category on a \emph{skew
  multicategory}~\cite{BL:multi} instead of a set: a Hilbert-style
calculus and a cut-free sequent calculus. These constructions
generalize the free construction on a set and do not generally entail
the left-normality condition.

As explained by Shulman~\cite[Rev.~49]{ncatlab:cloc}, prounital closed
categories are the natural notion of category with internal hom as the
only required connective (when we do not have/do not want a monoidal
structure $\I$,$\ot$ in the category), since they form an essential,
in the sense of minimal, class of models for planar typed
lambda-calculus.  This is the case because, formally, they are
equivalent to \emph{closed
  multicategories}~\cite{Man:clocvc}. Multicategories are models of
deductive systems with only identity and composition as basic
operations, while closed multicategories are also able to model
implication.  Standard approaches to denotational semantics of typed
lambda-calculus, adapted to the planar case, would exclude prounital
closed categories as valid models. This is because these approaches
usually require the presence of a Cartesian monoidal (just monoidal in
the planar case) structure complementing the closed structure. We
believe there is no good reason to discard models not interpreting the
non-existing connectives $\I,\ot$ and prounital closed categories are
the right notion of categorical model for planar typed lambda
calculus.  Analogously, skew prounital closed categories are the
correct notion of category with skew internal hom as the only required
connective, since they are equivalent to \emph{closed skew
  multicategories}~\cite{BL:multi}, a skew variant of closed
multicategories.

It is worth mentioning that there is another way of weakening closed categories by simply dropping all references to the unit altogether, that is, by only asking for internal hom objects equipped with the extranatural transformation $L$ and pentagon equation (c5) described below.
These may be called \emph{non-unital closed categories,} or \emph{semi-closed categories} after Bourke \cite{Bou:skews}, and are of some interest in providing interpretations for planar lambda terms with no closed subterms (a condition analogous to that of \emph{bridgelessness} in graph theory \cite{Zei:theltf}).
We do not treat nonunital closed categories explicitly here, although we expect that our results may be adapted from the prounital to the nonunital case in a straightforward way.

We fully formalized the results presented in the paper in the
dependently typed programming language Agda. The formalization uses
Agda version 2.6.0. and it is available at
\url{https://github.com/niccoloveltri/skew-prounital-closed-cats}.

\section{Skew Prounital Closed Categories}
\label{sec:skew}

A \emph{skew prounital closed category}~\cite{UVZ:eilkr} is a category
$\C$ equipped with functors $J : \C \to \Set$ (the \emph{element set}
functor) and ${\lo} : \Cop \times \C \to \C$ (the \emph{internal hom}
functor) and (extra)natural transformations $j$, $i$, $L$ typed
\[
\begin{array}{c}
j_A \in J\, (A \lo A) \qquad
i_{A,B} : J\, A \to \C(A \lo B, B) \qquad 
L_{A,B,C} \in \C(B \lo C, (A \lo B) \lo (A \lo C)) 
\end{array}
\]
satisfying the following equations where we write $\plug : \C(A,B)
\times J\, A \to J\, B$ for $J$ as a left action, i.e., $f \plug e =
J\, f\, e$:
\begin{description}
  \item{\rm{(c1)}} $e = i_{A,A}\, e \plug j_A \in J\, A$ for $e \in J\, A$;    
  \item{\rm{(c2)}} $i_{A\lo A,A \lo C}\, j_A \comp L_{A,A,C} = \id_{A \lo C} \in \C(A \lo C, A \lo C)$;
  \item{\rm{(c3)}} $L_{A,B,B} \plug j_B = j_{A \lo B} \in J\, ((A \lo B) \lo (A \lo B))$;
  \item{\rm{(c4)}} $i_{A,B}\, e \lo C = ((A \lo B) \lo i_{A,C}\, e) \comp L_{A,B,C} \in \C(B \lo C, (A \lo B) \lo C)$ for $e \in J\, A$;
  \item{\rm{(c5)}} $(B \lo C) \lo L_{A,B,D} \comp L_{B,C,D} = L_{A,B,C} \lo ((A \lo B) \lo (A \lo D)) \comp L_{A\lo B,A\lo C ,A\lo D} \comp L_{A,C,D} \linebreak \in \C(C\lo D,(B \lo C) \lo ((A \lo B) \lo (A \lo D)))$.
\end{description}

We typically write $\C(\n,B)$ for $J\,B$. Let $S$ be an
\emph{optional object}, i.e., $S$ is either nothing (denoted $S = \n$) or
it is an object of $\C$. We define $\C(S,B)$ as $J\, B$ if $S = \n$
and as $\C(A,B)$ if $S = A$. The use of this ``enhanced'' notion of
homset with an optional object as domain allows the unification of the
two notions of composition $\comp$ and $\plug$. We overload the
composition symbol $\circ$: given $f \in \C(S,B)$ and $g \in \C(B,C)$,
we write $g \circ f \in \C(S,C)$, which is equal to $g \plug f$ when
$S = \n$ and it is equal to the usual composition of maps $g \comp f$
when $S = A$. With the new notation, the types of structural laws $j$ and $i$ become
\[
j_A \in \C(\n,A \lo A) \qquad
i_{A,B} : \C(\n, A) \to \C(A \lo B, B)
\]

We note that it is not strictly necessary to require that $\lo$ is a
functor $\Cop \times \C \to \C$. It suffices to require that
$A \lo {} : \C \to \C$ is a functor for every $A$, since
the functorial action ${} \lo B$  
can for any $B$
be defined from the rest of the structure as $f
\lo B = i_{A\lo A',A\lo B}\, ((A \lo f) \comp j_A) \comp  L_{A,A',B} $ for $f : A
\to
A'$ and proved to preserve identity and composition and be natural in $B$ 
from the equations governing it.  
Under such an alternative definition of skew prounital
closed category, the equations (c4)--(c5) above have to be suitably
adjusted and an equation for bifunctoriality of $\lo$ added.

Skew prounital closed categories differ from Shulman's 
prounital closed categories in that the derivable map
\begin{equation}\label{eq:ln}
\begin{array}{l}
  \jY_{A,B} : \C(A,B) \to \C(\n,A \lo B) \\
  \jY_{A,B}\;f = (A \lo f) \comp j_A
\end{array}
\end{equation}
is not required to be a natural isomorphism. A skew prounital closed
category in which $\jY$ is invertible is called \emph{left-normal}.

We should note that Shulman's prounital closed categories, although
more normal than skew prounital closed categories, are nonetheless
partially skew. One could also require \emph{right-normality} and
\emph{associa\-tive-normality}, corresponding to invertibility of
certain derivable maps $\iY$ and $\LY$~\cite{UVZ:eilkr}. Eilenberg and Kelly's closed
categories are partially skew in that they are not associative-normal.

\begin{example}
  A simple example of a prounital closed category (adapted from de
  Schipper~\cite{dS:symcc}) is obtained by taking a suitable full
  subcategory of the skeletal version $\mathbb{F}$ of the Cartesian
  monoidal closed category of finite sets and functions (i.e., exactly
  one set of each finite cardinality). Namely, we keep only
  cardinalities from $M \subseteq \Nat$ given inductively by:
  $3 \in M$ and $n^m \in M$ for all $m, n \in M$. Now $1 \notin M$ and
  $3 \times 3 = 9 \notin M$, so we have lost the original unit
  $\I = 1$ and the tensor $p \ot m = p \times m$, but we still have
  the internal hom given by $m \lo n = n^m$. No other candidate unit
  or tensor can work since we need to have $m \cong \I \lo m$ and
  $\C(p \ot m, n) \cong \C(p, m \lo n)$.  Nevertheless, this category
  is prounital with the $J$ functor given by the composition of
  inclusions $M \hookrightarrow \mathbb{F} \hookrightarrow \Set$, and
  $j_m$ and $i_{m,n}$ defined in the evident way.  This category is
  left-normal, but neither right-normal nor associative-normal.

  To skew a closed category, one can use any left-strong monad on it
  \cite{UVZ:eilkr}; the same construction works for a prounital closed
  category (the concept of left-strength of a monad has to be adjusted
  to this setting).  We consider the reader monad given by $T m = m^k$
  for some fixed $k \in M$. The Kleisli category is skew prounital
  closed with the internal hom defined by
  $m \lo^T n = m \lo T n = n^{k\times m}$.  This category is neither
  left-normal nor right-normal or associative-normal.

  Alternatively, we can begin with $\mathbb{F}$ and take the full
  subcategory corresponding to $M = \Nat \setminus \{0,1\}$. This time
  we get a prounital monoidal closed category. We can skew it as
  before and we still get a skew prounital nonmonoidal closed category
  since the Kleisli construction destroys the tensor.
\end{example}

A \emph{strict prounital closed functor} between skew prounital closed
categories $\C$ and $\D$ consists of a functor $F : \C \to \D$ such
that $F\,(A \lo B) = F\, A \lo F\,B$, $F\,(f \lo g) = F\, f \lo F\,g$, and the structural laws $j$, $i$ and
$L$ are preserved on the nose. In particular, the functor $F$ is asked to
map $\C(S,B)$ to $\D(F\,S,F\,B)$, with $F\,S = \n$ if $S = \n$ and
$F\,S = F\,A$ if $S = A$, and preserve the enhanced notion of
composition $\comp : \C(B,C)\times \C(S,B) \to \C(S,C)$. 
Skew prounital closed categories and strict prounital closed functors
between them form a category.

\section{The Free Skew Prounital Closed Category on a Set}
\label{sec:free}

We now look at different presentations of the free skew prounital
closed category on a set of generating objects.  Let us first make
explicit the definition of this free construction.

The \emph{free} skew prounital closed category on a set $\Var$ is a
skew prounital closed category $\FskPCl(\Var)$ equipped with an
inclusion $\iota: \Var \to \FSkPCl(\Var)$, interpreting elements of
$\Var$ as objects of $\FSkPCl(\Var)$. For any other skew prounital
closed category $\C$ with a function $G : \Var \to \C$, there must
exist a unique strict prounital closed functor
$\bar{G} : \FSkPCl(\Var) \to \C$ compatible with $\iota$.

The existence of the free skew prounital closed category
$\FSkPCl(\Var)$ entails the existence of a left adjoint to the
forgetful functor between the category of skew prounital closed
categories and strict prounital closed functors and the category of
sets and functions.


\subsection{Categorical Calculus}
\label{sec:catcalc}

The first presentation consists of a deductive system that we call 
the \emph{categorical calculus} since it is directly derived from the definition of skew prounital closed category. (We could also think of it as a Hilbert-style calculus of sorts, or under the Curry-Howard correspondence, a combinatory logic.) 
Objects are \emph{formulae} inductively generated as follows: a
formula is either an element $X$ of $\Var$ (an \emph{atomic} formula) or
of the form $A \lo B$, where $A,B$ are formulae. We write $\Tm$ for the set
of formulae.

Maps between an optional formula $S$ and a formula $C$ are
\emph{derivations} of the sequent $S \tto C$, inductively generated by
the following inference rules:
\begin{equation}\label{eq:catcalc}\small
\renewcommand{\arraystretch}{1.6}
\begin{array}{c@{\quad \quad}c@{\quad \quad}c}
\infer[\id]{A \tto A}{}
&
\infer[\dcomp]{S \tto C}{S \tto B & B \tto C}
&
\infer[\lo]{A \lo B \tto C \lo D}{C \tto A & B \tto D}
\\
\infer[j]{ \tto A \lo A}{}
&
\infer[i]{A \lo B \tto B}{\tto A}
&
\infer[L]{B \lo C \tto (A \lo B) \lo (A \lo C)}{}
\end{array}
\end{equation}

Derivations are \emph{identified} up to a congruence relation
$\doteq$ that is inductively generated by the following pairs of
derivations:
\begin{equation}\label{eq:doteq}\small
\begin{array}{@{\quad}l@{\quad}c}
\text{(category laws)} &
\id \comp f \doteq f
\qquad
f \doteq f \comp \id
\qquad
(f \comp g) \comp h \doteq f \comp (g \comp h)
\\[5pt]
\text{(${\lo}$ functorial)} &
\id \lo \id \doteq \id
\qquad 
(f \comp h) \lo (k \comp g) \doteq (h \lo k) \comp (f \lo g)
\\[5pt]
&
f \lo \id \comp j \doteq \id \lo f \comp j
\\
\text{($j,i,L$ (extra)nat. trans.)} &
g \comp i(e) \comp h \lo \id \doteq i(h \comp e) \comp \id \lo g
\\
&
(f \lo g) \lo (\id \lo h) \comp L \doteq \id \lo (f \lo \id) \comp L \comp g \lo h
\\[5pt]
&
i (e) \comp j \doteq e
\qquad
i (j) \comp L \doteq \id
\\
(\textrm{c1-c5}) &
L \comp j \doteq j
\qquad
\id \lo i (e) \comp L \doteq i (e) \lo \id
\\
&
\id \lo L \comp L \doteq L \lo \id \comp L \comp L
\end{array}
\end{equation}
In the term notation for derivations, we write $g \comp f$ for
$\dcomp\, f\, g$ to agree with the standard categorical notation.

The categorical calculus defines the free skew prounital closed
category on $\Var$ in a straightforward way. Given another skew
prounital closed category $\C$ with function $G : \Var \to \C$, we can
easily define mappings $\bar{G}_0 : \Fma \to \C_0$ and $\bar{G}_1 : S
\tto C \to \C(\bar{G}_0(S),\bar{G}_0(C))$ by induction. These specify
a strict prounital closed functor, in fact the only existing one
satisfying $\bar{G}_0(X) = G(X)$.

\subsection{Cut-Free Sequent Calculus}
\label{sec:seqcalc}

The second presentation of $\FSkPCl(\Var)$ is a sequent
calculus. Sequents are triples of the form $\stseq S \Gamma C$. The
succedent $C$ is a formula in $\Fma$. The antecedent is split in two
parts: the \emph{stoup} $S$ is an optional formula, i.e. it is either
empty or it is a single formula; the \emph{context} $\Gamma$ is a list
of formulae.

Derivations in the sequent calculus are inductively generated by the
following inference rules:
\begin{equation}\label{eq:seqcalc}\small
\renewcommand{\arraystretch}{1.6}
\begin{array}{c@{\quad \quad}c}
\infer[\uf]{\stseq{\n} {A, \Gamma}{C}}{
  \stseq{A}{\Gamma} C
}
&
\infer[\loR]{\stseq{S}{\Gamma}{A \lo B}}{
  \stseq{S}{\Gamma,A}{B}
}
\\[5pt]
\infer[\ax]{\stseq A {~} A}{
}
&
\infer[\loL]{\stseq{A \lo B}{\Gamma,\Delta} C}{
  \stseq{\n}{\Gamma}{A} & \stseq{B}{\Delta} C
}
\end{array}
\end{equation}
($\uf$ for `passivate', $\mathsf{L}$, $\mathsf{R}$ for introduction
on the left (in the stoup) resp.\ right)
and identified up to the congruence $\circeq$ induced by the equations:
\begin{equation}\label{eq:circeq}\small
\begin{array}{l@{\quad}c@{\quad}l}
  \textrm{($\eta$-conversion)} &
  \ax_{A \lo B} \circeq \loR\;(\loL\;(\uf\;\ax_A,\ax_B)) &
  \\[9pt]
  \multicolumn{3}{l}{\textrm{(commutative conversions)}} \\
  & \uf\;(\loR\;f) \circeq \loR\;(\uf\;f)
  & (\text{for } f : \stseq{A'}{\Gamma,A}{B})
  \\
  & \loL\;(f , \loR\;g) \circeq \loR\;(\loL\;(f,g))
  & (\text{for } f : \stseq{\n}{\Gamma}{A'}, \; g : \stseq{B'}{\Delta,A}B)
\end{array}
\end{equation}

There are no primitive cut rules in this sequent calculus, but two forms of
cut are admissible:
\begin{equation}\label{eq:cut}\small
\infer[\scut]{\stseq S {\Gamma, \Delta} C}{
  \stseq S {\Gamma} A
  &
  \stseq{A} {\Delta} C
}
\qquad
\infer[\ccut]{\stseq S {\Delta_0, \Gamma, \Delta_1} C}{
  \stseq {\n} {\Gamma} A
  &
  \stseq {S} {\Delta_0, A, \Delta_1} C
}
\end{equation}

Notice that the left rule $\loL$ acts only on the implication $A \lo
B$ in the stoup. Another left rule $\loC$ acting on implication in the
passive context is derivable from cut:
\begin{equation}\label{eq:loC}
\small
\proofbox{
  \infer[\loC]{\stseq{S}{\Delta_0,A \lo B,\Gamma,\Delta_1}{C}}{
  \stseqname{\n}{\Gamma}{f}{A}
  &
  \stseqname{S}{\Delta_0,B,\Delta_1}{g}{C}
  }
  }
 = 
\proofbox{
  \infer[\ccut]{\stseq{S}{\Delta_0,A \lo B,\Gamma,\Delta_1}{C}}{
  \infer[\uf]{\stseq{\n}{A \lo B , \Gamma}{B}}{
    \infer[\loL]{\stseq{A \lo B}{\Gamma}{B}}{
      \stseqname{\n}{\Gamma}{f}{A}
      &
      \infer[\ax]{\stseq{B}{~}{B}}{}
    }
  }
  &
  \stseqname{S}{\Delta_0,B,\Delta_1}{g}{C}
  }
}
\end{equation}

\paragraph{Soundness.}
Sequent calculus derivations can be turned into categorical calculus
derivations using a function $\sound : (\stseq{S}{\Gamma}{C}) \to (S \tto
\asem{\Gamma}{C})$, where the formula $\asem{\Gamma}{C}$ is inductively
defined as
\[
\asem{~} C = C
\qquad\qquad
\asem{A , \Gamma} C = A \lo \asem{\Gamma}C
\]

Given $f : \stseq{S}{\Gamma,A}{B}$, define $\sound(\loR\;f)$ simply as $\sound(f)$. Given $f : \stseq{A}{\Gamma}{C}$, define $\sound(\uf\;f)$ as
\[\small
\infer={ \tto \asem{A ,\Gamma} C}{
  \infer[\dcomp]{ \tto A \lo \asem{\Gamma} C}{
    \infer[j]{\tto A \lo A}{}
    &
    \infer[\lo]{A \lo A \tto A \lo \asem{\Gamma} C}{
      \infer[\id]{A \tto A}{}
      &
      A \stackrel{\sound(f)}{\tto} \asem{\Gamma}C
    }
  }
}
\]
The double-line rule corresponds to the application of the 
equality $\asem{A,\Delta}C = A \lo \asem{\Delta}C$.
Given \linebreak $f : \stseq{\n}{\Gamma}{A}$ and $g : \stseq{B}{\Delta}{C}$, define $\sound(\loL(f,g))$ as
\[\small
\infer[\dcomp]{A \lo B \tto \asem{\Gamma ,\Delta} C}{
  \infer[{\lo}]{A \lo B \tto A \lo \asem{\Delta} C}{
    \infer[\id]{A \tto A}{}
    &
    B \stackrel{\sound(g)}{\tto} \asem{\Delta} C
  }
  &
  \infer[\dcomp]{A \lo \asem{\Delta} C \tto \asem{\Gamma, \Delta} C}{
    \infer[\Lstar]{A \lo \asem{\Delta} C \tto \asem{\Gamma} A \lo \asem{\Gamma,\Delta} C}{}
    &
    \infer[i]{\asem{\Gamma} A \lo \asem{\Gamma,\Delta} C \tto \asem{\Gamma,\Delta}C}{
      \stackrel{\sound(f)}{\tto} \asem{\Gamma}A
    }
  }
}
\]
where the operation $\Lstar$, defined by induction on $\Gamma$, 
performs iterated applications of the structural law $L$.  The
function $\sound$ is well-defined, in the sense that it sends
$\circeq$-equivalent derivations to $\doteq$-related derivations.

\paragraph{Completeness.}
Derivations in the categorical calculus can be turned into sequent
calculus derivations via a function $\cmplt : (S \tto \asem{\Gamma}C)
\to (\stseq{S}{\Gamma}{C})$ . The $\ax$ rule models the identity map,
while sequential composition is interpreted using $\scut$.
Functoriality of $\lo$ is modelled using $\loR$ and $\loL$.
The function $\cmplt$ sends the structural laws $j,i,L$ of skew
prounital closed categories to the following derivations in the
sequent calculus:
\[\small
\begin{array}{lc}
  (j) &
\proofbox{\infer[\loR]{\stseq{\n}{~}{A \lo A}}{
  \infer[\uf]{\stseq{\n}{A}{A}}{
    \infer[\ax]{\stseq{A}{~}{A}}{}
  }
}}
\\[30pt]
  (i) &
\proofbox{\infer[\loL]{\stseq{A \lo B}{~}{B}}{
  \stseq{\n}{~}{A}
  &
  \infer[\ax]{\stseq{B}{~}{B}}{}
}}
\end{array}
\qquad
\begin{array}{cr}
  \proofbox{
    \infer[\loR]{\stseq{B \lo C}{~}{(A \lo B) \lo (A \lo C)}}{
    \infer[\loR]{\stseq{B \lo C}{A \lo B}{A \lo C}}{
      \infer[\loL]{\stseq{B \lo C}{A \lo B,A}{C}}{
        \infer[\uf]{\stseq{\n}{A \lo B,A}{B}}{
          \infer[\loL]{\stseq{A \lo B}{A}{B}}{
            \infer[\uf]{\stseq{\n}{A}{A}}{
              \infer[\ax]{\stseq{A}{~}{A}}{}
            }
            &
            \infer[\ax]{\stseq{B}{~}{B}}{}
          }
        }
        &
        \infer[\ax]{\stseq{C}{~}{C}}{}
      }
    }
    }
  }
  &
  (L)
\end{array}
\]
The function $\cmplt$ is well-defined, in the sense that it sends
$\doteq$-equivalent derivations to $\circeq$-related
derivations. Moreover it is possible to prove that $\cmplt$ is the
inverse of $\sound$ (up to the equivalence relations $\doteq$ and
$\circeq$). This shows that the sequent calculus is a presentation of
the free skew prounital closed category $\FskPCl(\Var)$.

\subsection{Natural Deduction Calculus}
\label{sec:natded}

The third presentation of $\FSkPCl(\Var)$ is a natural deduction calculus. 
Sequents
are triples $\stseqnd S \Gamma C$ as in the sequent calculus of
Section~\ref{sec:seqcalc}, but the left rule $\loL$ for $\lo$ is replaced by
the elimination rule $\loe$. 
\[\small
\begin{array}{c@{\quad \quad}c}
\infer[\uf]{\stseqnd{\n} {A, \Gamma}{C}}{
  \stseqnd{A}{\Gamma} C
}
&
\infer[\loi]{\stseqnd{S}{\Gamma}{A \lo B}}{
  \stseqnd{S}{\Gamma,A} B
}
\\[5pt]
\infer[\ax]{\stseqnd{A}{~} A}{
}
&
\infer[\loe]{\stseqnd{S}{\Gamma,\Delta} B}{
  \stseqnd{S}{\Gamma}{A \lo B} &
  \stseqnd{\n}{\Delta} A
}
\end{array}
\]
The two cut rules in (\ref{eq:cut}) are admissible also in the
natural deduction calculus. Derivations are identified by the congruence relation
$\doteqdot$, a skew ordered variant of the usual
$\beta\eta$-equivalence of simply typed lambda-calculus, induced by
the equations
\begin{equation}\label{eq:doteqdot}\small
\begin{array}{l@{\quad}c@{\qquad}l}
  \textrm{($\beta$-conversion)} &
  \loe \;(\loi \;f,g) \doteqdot \ccut\;(g,f)
  & (\text{for } f : \stseqnd{S}{\Gamma,A}{B}, \;g : \stseqnd{\n}{\Delta}{A})  
  \\[9pt]
  \textrm{($\eta$-conversion)} &
  f \doteqdot \loi\;(\loe\;(f, \uf\;\ax))
  & (\text{for } f : \stseqnd{S}{\Gamma}{A \lo B})  
  \\[9pt]
    \multicolumn{3}{l}{\textrm{(commutative conversions)}} \\
  & \uf\;(\loi\;f) \doteqdot \loi\;(\uf\;f)
  & (\text{for } f : \stseqnd{A'}{\Gamma,A}{B})
  \\
  &  \uf\;(\loe\;(f,g)) \doteqdot \loe\;(\uf\;f,g)
  & (\text{for } f : \stseqnd{A'}{\Gamma}{A \lo B}, \; g : \stseqnd{\n}{\Delta}{A})
\end{array}
\end{equation}

This natural deduction calculus corresponds to a variant of the planar
fragment of linear typed lambda-calculus~\cite{Abr:temper,Zei:theltf}. The formulae correspond to types. The derivations correspond to lambda terms in which all free and bound variables are used exactly once and in the order of their declaration. 
The equations axiomatize the appropriate variant of $\beta\eta$-equivalence.



It is possible to prove that the natural deduction calculus is
equivalent to the sequent calculus (up to the equivalences of
derivations $\circeq$ and $\doteqdot$), and it is therefore a
presentation of $\FSkPCl(\At)$. We do not follow this strategy
here. Instead we construct reduction-free normalization procedures for
the sequent calculus and the natural deduction calculus. The
procedures target two calculi of normal forms: a focused subsystem of
the sequent calculus and a calculus of $\beta\eta$-long normal forms
wrt. $\doteqdot$. By showing that the calculi of normal forms are
equivalent, we conclude that the sequent calculus and the natural
deduction calculus are also equivalent up to the equivalences of
derivations $\circeq$ and $\doteqdot$.

\subsection{Focused Sequent Calculus Derivations}
\label{sec:focus}

The congruence relation $\circeq$ on sequent calculus derivations can be considered as a term rewrite
system, by directing every equation in (\ref{eq:circeq}) from left to
right. The resulting rewrite system is weakly confluent and strongly
normalizing, hence confluent with unique normal forms.

Derivations in normal form wrt.\ $\circeq$ can be described by a
suitable \emph{focused} subcalculus of the full sequent calculus,
following the paradigm introduced by
Andreoli~\cite{And:logpfp}. Derivations in this subcalculus are
inductively generated by the following inference rules:
\begin{equation} \label{eq:foccalc}\small
\renewcommand{\arraystretch}{2.5}
\begin{array}{c@{\quad}|@{\quad}c@{\quad}|@{\quad}c}
\infer[\loR]{\stseqI{S}{\Gamma}{A \lo B}}{
  \stseqI{S}{\Gamma,A}{B}
}
&
\infer[\uf]{\stseqP{\n} {A, \Gamma}{C}}{
  \stseqP{A}{\Gamma} C
}
&
\infer[\ax]{\stseqF A {~} A}{
}
\\[5pt]
\infer[\mathsf{P2I}]{\stseqI{S}{\Gamma}{X}}{
  \stseqP{S}{\Gamma}{X}
}
&
\infer[\mathsf{F2P}]{\stseqP{A}{\Gamma}{C}}{
  \stseqF{A}{\Gamma}{C}
}
&
\infer[\loL]{\stseqF{A \lo B}{\Gamma,\Delta} C}{
  \stseqI{\n}{\Gamma}{A} & \stseqF{B}{\Delta} C
}
\end{array}
\end{equation}
This is a sequent calculus with an additional phase annotation on
sequents, for controlling root-first proof search. In phase \textsf{I} (for \emph{inversion}), sequents have
the form $\stseqI{S}{\Gamma}{C}$, where $S$ is a general stoup and $C$
is a general formula. During this phase, we eagerly apply the
invertible rule $\loR$ until the succedent is reduced to an atomic
formula. In phase \textsf{P} (for \emph{passivation}), 
we have the opportunity of applying the
$\uf$ rule and can only go to the last phase \textsf{F} (for \emph{focusing})
when the stoup has become a formula. During this phase we
can finish the derivation using $\ax$, which is now restricted to
atomic formulae, or apply the $\loL$ rule. If we apply the $\loL$ rule,
we are thrown back to the $\textsf{I}$ phase in the first premise. One can observe that,
in any $\textsf{I}$-derivation, the succedent of any $\textsf{P}$- or $\textsf{F}$-sequent must actually be an atom, but the 
generality of allowing any formula in the succedent in these phases (which we can have in derivations of $\textsf{P}$- or $\textsf{F}$-sequents) will be useful for us shortly in the discussion of hereditary substitutions below 
and also in Section~\ref{sec:norm} where we will relate focused sequent calculus derivations to normal natural deduction calculus derivations.

\paragraph{Focusing.}
The focused rules define a sound and complete root-first proof search strategy
for the cut-free sequent calculus of Section~\ref{sec:seqcalc}. 
Soundness of the focused calculus is evident: focused derivations can be
embedded into sequent calculus derivations via functions $\emb_k : (S
\mid \Gamma \longrightarrow_k C) \to (\stseq{S}{\Gamma}{C})$ for all phases $k \in
\{\mathsf{I},\mathsf{P},\mathsf{F} \}$ that just erase all
phase annotations and uses of the rules $\mathsf{P2I}$ and $\mathsf{F2P}$.

By the normalization property of the rewrite system associated to
$\circeq$, we know that the focused calculus is also complete. This
can also be established by constructing a reduction-free normalization
function $\focus : (\stseq{S}{\Gamma}{C}) \to (\stseqI{S}{\Gamma}{C})$
sending each derivation in the sequent calculus to a canonical
representative of its $\circeq$-equivalence class in the focused
calculus. This means in particular that $\focus$ maps
$\circeq$-related derivations to equal focused derivations. For the
definition of $\focus$, we show that general $\uf$, $\loL$ and $\ax$
rules are admissible in phase \textsf{I}:
\[\small
\infer[\uf^{\mathsf{I}}]{\stseqI{\n} {A, \Gamma}{C}}{
  \stseqI{A}{\Gamma} C
}
\qquad\qquad
\infer[\loL^{\mathsf{I}}]{\stseqI{A \lo B}{\Gamma,\Delta} C}{
  \stseqI{\n}{\Gamma}{A} & \stseqI{B}{\Delta} C
}
\qquad\qquad
\infer[\ax^{\mathsf{I}}]{\stseqI A {~} A}{
}
\]
This makes each sequent calculus inference rule matched by a focused
calculus rule. Then the normalization procedure $\focus$ can be easily
defined by induction on the input derivation. The function $\focus$ is
the inverse of $\emb_{\mathsf{I}}$ up to $\circeq$: given a sequent
calculus derivation $f : \stseq{S}{\Gamma}{C}$, we have
$\emb_{\mathsf{I}}\;(\focus\;f) \circeq f$; given a focused derivation
$f : \stseqI{S}{\Gamma}{C}$, we have $\focus\;(\emb_{\mathsf{I}}\;f) = f$.

The focused calculus solves the \emph{coherence problem} for
skew prounital closed categories, in the sense of giving an explicit
characterization of the homsets of $\FSkPCl(\Var)$.  It also solves
two related algorithmic problems effectively:
\begin{itemize}
\item Duplicate-free enumeration of all maps $S \tto C$ in the form of
  representatives of $\doteq$-equivalence classes of categorical
  calculus derivations: For this, find all focused derivations of
  $\stseqI S {~} C$, which is solvable by exhaustive proof search,
  which terminates, and translate them to the categorical calculus
  derivations.
\item Finding whether two given maps of type $S \tto C$, presented as
  categorical calculus derivations, are equal, i.e., $\doteq$-related
  as derivations: For this, translate them to focused derivations of
  $\stseqI S {~} C$ and check whether they are equal, which is
  decidable.
\end{itemize}

Monoidal and nonmonoidal closed categories, and prounital closed
categories likewise, admit no simple coherence theorem like Mac Lane's
for monoidal categories \cite{ML:natac} (depending on an easy
condition on the domain and codomain, no maps or just one in a
homset). Enumeration of (presentations of) maps and equality checking
are nontrivial \cite{KML:cohcc,Lap:cohncc,Min:cloctp,Sza:algp}. In our
focused calculus, the sequent
$\stseqI{(X \lo Y) \lo (X \lo Z)}{X \lo Y, X \lo X, X}{Z}$ has two
distinct focused derivations (we learned this example from Anupam
Das).

\paragraph{Hereditary Substitutions.}
Focused sequent calculus derivations can also be used as normal forms 
for the natural deduction
calculus of Section~\ref{sec:natded}. 
We show this by describing a reduction-free normalization procedure 
that is
typically called \emph{normalization by hereditary
  substitution}~\cite{WCPW:conlfp,KA:hersst}. Normal forms for this
procedure (at least in the case of simply-typed lambda-calculus) 
are typically defined in terms of a suitable spine calculus,
but we have defined our focused sequent calculus 
liberally enough to serve this purpose.

The focused calculus defines a sound and complete root-first proof search
strategy for the natural deduction calculus of
Section~\ref{sec:natded}. Focused derivations can easily
be embedded into natural deduction derivations:
there are functions $\emb^{\mathsf{nd}}_k : (S \mid
\Gamma \longrightarrow_k C) \to (\stseqnd{S}{\Gamma}{C})$ for all $k \in
\{\mathsf{I},\mathsf{P},\mathsf{F} \}$.

Similar to focusing completeness, 
normalization by hereditary substitution is also specified in
two steps. First we need to show that $\ax$ and $\loe$
rules are admissible in phase \textsf{I}:
\[\small
\infer[\ax^{\mathsf{I}}]{\stseqI A {~} A}{
}
\qquad\qquad
\infer[\loe^{\mathsf{I}}]{\stseqI{S}{\Gamma,\Delta} B}{
  \stseqI{S}{\Gamma}{A \lo B} & \stseqI{\n}{\Delta} A
}
\]
(We already know that a general $\uf$ rule is admissible in phase
\textsf{I}\footnote{We also already know from focusing that $\ax^{\mathsf{I}}$ is admissible, but it is defined in terms of $\loL^{\mathsf{I}}$. For normalization by hereditary substitutions, we define $\ax^{\mathsf{I}}$ differently, avoiding the use of $\loL^{\mathsf{I}}$ since $\loL$ is not a primitive rule of the natural deduction calculus.}.)
Focused derivations in phase $\textsf{I}$ 
should correspond to normal forms, i.e., 
canonical representatives of $\doteqdot$-equivalence classes. In
particular, they should not contain any redex. This forces the rule
$\loe^{\mathsf{I}}$ to be simultaneously defined with 3 pairs of substitution rules (i.e., cut rules) in the focused calculus, one for each phase $k
\in \{ \mathsf{I},\mathsf{P},\mathsf{F}\}$:
\[\small
\infer[\scut^k]{\stseqk S {\Gamma, \Delta} C}{
  \stseqk S {\Gamma} A
  &
  \stseqk{A} {\Delta} C
}
\qquad
\infer[\ccut^k]{\stseqk S {\Delta_0, \Gamma, \Delta_1} C}{
  \stseqk {\n} {\Gamma} A
  &
  \stseqk {S} {\Delta_0, A, \Delta_1} C
}
\]
The rule $\loe^{\mathsf{I}}$ can then be defined as dictated by the
$\beta$-conversion equation in the definition of $\doteqdot$ as
\[\small
\infer[\loe^{\mathsf{I}}]{\stseqI{S}{\Gamma,\Delta} B}{
  \infer[\loi]{\stseqI{S}{\Gamma}{A \lo B}}{
    \stseqI[f]{S}{\Gamma,A}{B}
  }
  & \stseqI[g]{\n}{\Delta} A
}
\quad = \quad
\infer[\ccut^{\mathsf{I}}]{\stseqI{S}{\Gamma,\Delta} B}{
  \stseqI[g]{\n}{\Delta} A
  &
  \stseqI[f]{S}{\Gamma,A}{B}
}
\]
Notice that the first premise is forced to be of the form $\loi\;f$.
This simultaneous substitution in canonical forms and reduction of
redexes that appear from substitution is the main idea behind
hereditary substitutions.  We can then construct a normalization
function $\hered : (\stseqnd{S}{\Gamma}{C}) \to (\stseqI{S}{\Gamma}{C})$
sending each primitive rule of the natural deduction calculus 
to its admissible
counterpart in the focused calculus. In particular, the function
$\hered$ maps each natural deduction derivation 
to its normal form 
as rendered in the focused calculus (which is our spine calculus). 

The function $\hered$ is the inverse of
$\emb^{\mathsf{nd}}_{\mathsf{I}}$ up to $\doteqdot$: given a 
natural deduction derivation \linebreak $f : \stseqnd{S}{\Gamma}{C}$, we have
$\emb^{\mathsf{nd}}_{\mathsf{I}}\;(\hered\;f) \doteqdot f$; given 
a focused derivation
$f : \stseqI{S}{\Gamma}{C}$, we get
$\hered\;(\emb^{\mathsf{nd}}_{\mathsf{I}}\;f) = f$.

\subsection{Normal Natural Deduction Derivations}
\label{sec:norm}

The congruence relation $\doteqdot$ on natural deduction calculus
derivations also has normal forms, which correspond precisely to
$\beta\eta$-long normal forms in the familiar terminology of
lambda-calculus.

Derivations in normal form wrt.\ $\doteqdot$ can be described by a
suitable subcalculus of the full natural deduction
calculus. Derivations in this subcalculus are inductively generated by
the following inference rules:
\[\small
\renewcommand{\arraystretch}{2.5}
\begin{array}{c@{\quad}|@{\quad}c@{\quad}|@{\quad}c}
\infer[\loi]{\nf{S}{\Gamma}{A \lo B}}{
  \nf{S}{\Gamma,A} B
}
&
\infer[\uf]{\ndp{\n} {A, \Gamma}{C}}{
  \ndp{A}{\Gamma} C
}
&
\infer[\ax]{\ne{A}{~} A}{
}
\\[5pt]
\infer[\mathsf{p2nf}]{\nf{S}{\Gamma}{X}}{
  \ndp{S}{\Gamma}{X}
}
&
\infer[\mathsf{ne2p}]{\ndp{A}{\Gamma}{C}}{
  \ne{A}{\Gamma}{C}
}
&
\infer[\loe]{\ne{A'}{\Gamma,\Delta} B}{
  \ne{A'}{\Gamma}{A \lo B} & \nf{\n}{\Delta} A
}
\end{array}
\]
Derivations are organized in an introduction phase and in an
elimination phase~\cite{Pra:natd}. In lambda-calculus jargon, we refer
to derivations in these phases as (pure) \emph{normal forms} and
\emph{neutrals}. Normal forms are derivations of sequents of the
general form $\nf{S}{\Gamma}{C}$. Similarly to the case of
simply-typed lambda-calculus, a normal form is an iteration of
$\lambda$-abstraction on a neutral term of an atomic type. Neutrals
are derivations of sequents of the form $\ne{A}{\Gamma}{C}$ where the
stoup is required to be a formula. Intuitively, they correspond to
terms which are stuck for $\doteqdot$-conversion. A neutral is either
a variable (declared in the stoup, in our case) or a function
application that cannot compute due to the presence of another neutral
in the function position. Due to the skew aspect of our natural
deduction calculus, we also add an intermediate third phase
\textsf{p}, with sequents of the form $\ndp{A}{\Gamma}{C}$, in
which we have the choice of applying the structural rule $\uf$. This
is analogous to the passivation phase of the focused sequent calculus
of Section~\ref{sec:focus}. 

The normal natural deduction calculus defines a root-first proof
search strategy for the natural deduction calculus. This procedure is
sound. Normal forms can easily be embedded into natural deduction
derivations: there are functions $\emb^{\mathsf{nd}}_k : (S \mid \Gamma
\longrightarrow_k C) \to (\stseqnd{S}{\Gamma}{C})$ for all $k \in
\{\mathsf{nf},\mathsf{p},\mathsf{ne} \}$.

\paragraph{Normalization by Evaluation.}
The completeness of the normal natural deduction calculus, implying
that normal natural deduction derivations are indeed
$\beta\eta$-long normal forms, is proved via \emph{normalization by
  evaluation}~\cite{BS:inveft,AHS:catrrf}.  This is a reduction-free
procedure in which terms are first evaluated into a certain semantic
domain, and their evaluations are then reified back into normal forms.

We begin by constructing two discrete categories: $\Cxt$ and $\SCxt$.  The
category $\Cxt$ has lists of formulae as objects.
It has a strict monoidal
structure with the empty list as unit and concatenation of lists as
tensor. The category $\SCxt$ has objects of the form $S\mid\Gamma$,
with $S$ an optional formula and $\Gamma$ a list of formulae. 
It has a unit object $\n\mid ~$ (in which
both components are empty) and there exists an action of the monoidal
category $\Cxt$ on $\SCxt$: $(S \mid \Gamma) \cdot \Gamma' = S \mid
\Gamma,\Gamma'$. There is a functor $E : \Cxt \to \SCxt$,
sending each list $\Gamma$ to the pair $\n\mid\Gamma$.

We consider the two presheaf categories
$\Set^{\Cxt}$ and
$\Set^{\SCxt}$. The monoidal structure on $\Cxt$ lifts to the
\emph{Day convolution} monoidal closed structure on $\Set^{\Cxt}$:
\[
\begin{array}{c}
\I_\cxt \;\Gamma = (\Gamma = ()) 
\qquad
(P \ot_\cxt Q) \;\Gamma = 
\sum_{\Gamma_0,\Gamma_1} (\Gamma = \Gamma_0, \Gamma_1) \times P \;\Gamma_0 \times Q \;\Gamma_1
\\[6pt]
(Q \lo_\cxt P) \;\Gamma = \prod_{\Delta} \; Q \;\Delta \to P\;(\Gamma,\Delta)
\end{array}
\]
(Here and below $()$ denotes the empty list.)

The unit of $\SCxt$ lifts to a unit in $\Set^{\SCxt}$ given by
$\I_\scxt \;(S \mid \Gamma) = (\Gamma = ()) \times (S = \n)$.
The action of $\Cxt$ on $\SCxt$ lifts to an action of
$\Set^{\Cxt}$ on $\Set^{\SCxt}$:
\[\begin{array}{c}
(P \otimes_\scxt Q) \;(S \mid \Gamma) =
\sum_{\Gamma_0,\Gamma_1}  (\Gamma = \Gamma_0, \Gamma_1) \times P\;(S \mid \Gamma_0) \times Q \;\Gamma_1
\end{array}
\]
The functor $\otimes_\scxt Q$ has a right adjoint $Q \lo_\scxt$ given by:
\[
\begin{array}{c}
  (Q \lo_\scxt P)\;(S \mid \Gamma) = \prod_{\Delta} \; Q \;\Delta \to P\;(S \mid \Gamma,\Delta)
\end{array}
\]

The first step of normalization by evaluation is the interpretation of
syntactic constructs, i.e. formulae and natural deduction derivations, as semantic entities
in $\Set^{\SCxt}$. Formulae are modelled as presheaves over
$\SCxt$. Implication is modelled via the functor
$\lo_\scxt$. Notice the composition with the functor $E$, which is
needed for the interpretation to be well-defined. The interpretation
of an atomic formula $X$ on an object $S \mid \Gamma$ is the set of
normal forms of type $\nf S{\Gamma}{X}$.
\[
\dsem{X}\;(S \mid \Gamma) = \nf{S}{\Gamma}{X}
\qquad\qquad
\dsem{A \lo B}\;(S \mid \Gamma) = ((\dsem{A} \circ E) \lo_\scxt \dsem{B}) \;(S \mid \Gamma)
\]

Lists of formulae can be interpreted as presheaves over $\Cxt$.
\[
\dsem{~}\;\Delta = \I_\cxt \;\Delta
\qquad\qquad
\dsem{A,\Gamma}\;\Delta = ((\dsem{A} \circ E) \ot_\cxt \dsem{\Gamma})\;\Delta
\]
Finally, antecedents $S\mid\Gamma$ can be interpreted as presheaves
over $\SCxt$:
\[
\begin{array}{c}
\dsem{\n\mid\Gamma}\;(S\mid\Delta) 
= (\I_\scxt \otimes_\scxt \dsem{\Gamma}) \;(S\mid\Delta)
= (S= \n) \times \dsem{\Gamma}\;\Delta
\\[6pt]
\dsem{A\mid\Gamma}\;(S\mid\Delta) = (\dsem{A} \otimes_\scxt \dsem{\Gamma}) \;(S\mid\Delta)
\end{array}
\]

The next step of normalization by evaluation is the interpretation of a
derivation $f : \stseqnd{S}{\Gamma}{C}$ in the natural deduction calculus as a
natural transformation between presheaves $\dsem{S\mid\Gamma}$ and
$\dsem{C}$. Formally, we define an evaluation function by
induction on the input derivation:
\[
\eval : (\stseqnd{S}{\Gamma} C) \to \dsem{S \mid \Gamma}\;(S' \mid \Delta)
\to \dsem{C}\;(S' \mid \Delta)
\]

Subsequently, we extract a normal form from the evaluated term. The
reification procedure sends a semantic element in $\dsem{A}\;(S\mid
\Gamma)$ to a normal form in $\nf{S}{\Gamma}{A}$. The latter is
defined by mutual induction with a function reflecting neutrals
in $\ne{A}{\Gamma}{C}$ to semantic elements in $\dsem{C}\;(A \mid
\Gamma)$.
\[
\reflect : (\ne{A}{\Gamma}{C}) \to \dsem{C}\;(A \mid \Gamma)
\qquad\qquad
\reify: \dsem{A}\;(S \mid \Gamma) \to (\nf{S}{\Gamma}{A})
\]

Finally, a normalization procedure $\nbe : (\stseqnd{S}{\Gamma}{C}) \to
(\nf{S}{\Gamma}{C})$ is defined as follows. Apply $\eval$ to a given
derivation $f : \stseqnd{S}{\Gamma}{C}$ in the natural deduction calculus, obtaining a
natural transformation $\eval\;f$ between presheaves $\dsem{S \mid
  \Gamma}$ and $\dsem{C}$. Take the component of $\eval \;f$ at $S
\mid \Gamma$, which is a function of type $\dsem{S \mid \Gamma}\;(S \mid
\Gamma) \to \dsem{C}\;(S \mid \Gamma)$. By induction on $S$ and
$\Gamma$, it is possible to define a canonical element $\gamma :
\dsem{S \mid \Gamma}(S \mid \Gamma)$. This allows to obtain an element
$\eval\;f\;\gamma : \dsem{C}\;(S \mid \Gamma)$, which can finally be
reified into a normal form:
\[
\nbe\;f = \reify\;(\eval\;f\;\gamma)
\]
We formally verified that the function $\nbe$ is well-defined, i.e. it
sends $\doteqdot$-related derivations to the same normal form. Moreover,
$\nbe$ is the inverse up to $\doteqdot$ of the embedding 
 $\emb^{\mathsf{nd}}_{\mathsf{nf}}
: (\nf{S}{\Gamma}{C}) \linebreak \to (\stseqnd{S}{\Gamma}{C})$ of
normal forms into natural deduction derivations: given a 
natural deduction derivation $f : \stseqnd{S}{\Gamma}{C}$, we have
$\emb^{\mathsf{nd}}_{\mathsf{nf}}\;(\nbe\;f) \doteqdot f$; given a
normal form $f : \nf{S}{\Gamma}{C}$, we have
$\nbe\;(\emb^{\mathsf{nd}}_{\mathsf{nf}}\;f) = f$.

\paragraph{Comparing Normal Forms.}
We conclude this section by showing that focused sequent calculus
derivations and normal natural deduction derivations are
in one-to-one correspondence.  Notice that we have already established
a one-to-one correspondence indirectly: the correctness of
normalization by hereditary substitution implies that the set of
focused calculus derivations $\stseqI{S}{\Gamma}{C}$ is isomorphic to
the set of natural deduction derivations $\stseqnd{S}{\Gamma}{C}$
quotiented by the equivalence relation $\doteqdot$, which is
further isomorphic to the set of normal natural deduction derivations
$\nf{S}{\Gamma}{C}$ thanks to the correctness of normalization by
evaluation. The goal of this section is to provide a simple direct
comparison of the two classes of normal forms.

The crucial step of this comparison is the relation between neutrals
and derivations in phase \textsf{F}. This is because
normal forms in phase \textsf{nf} have the same primitive inference rules
derivations in phase \textsf{I}, and similarly for derivations of the
passivation phases of the two calculi. We simultaneously define six
translations back and forth between the three pairs of corresponding phases 
of the two calculi. We
only show the constructions of the translations $\netoF$ and $\Ftone$
between neutrals and derivations in phase \textsf{F}. The
functions $\nftoI$ and $\Itonf$ for translating between normal forms
and derivations in phase \textsf{I} are trivially
defined, similarly for the functions translating between the $\textsf{p}$ and
\textsf{P} phases. The definitions of translations $\netoF$ and
$\Ftone$ rely on two auxiliary functions $\netoF'$ and $\Ftone'$.
\[\arraycolsep=2pt \small
\begin{array}{ll}
  \multicolumn{2}{c}{
    \netoF' : (\ne{A}{\Gamma}{B}) \to (\stseqF{B}{\Delta}{C}) \to
  (\stseqF{A}{\Gamma,\Delta}{C})} \\
  \netoF'\;\ax & g = g \\
  \netoF'\;(\loe\;(f,a)) & g = \netoF' \;f \;(\loL\; (\nftoI\;a, g))
\end{array}
\]
\[\arraycolsep=2pt\small
\begin{array}{ll}
  \multicolumn{2}{c}{
    \Ftone' : (\ne{A}{\Gamma}{B}) \to (\stseqF{B}{\Delta}{C}) \to
  (\ne{A}{\Gamma,\Delta}{C})} \\
  \Ftone'\;f \;\ax & = f \\
  \Ftone'\;f \;(\loL\;(a,g)) &= \Ftone' \;(\loe\; (f ,\Itonf\;a))\;g
\end{array}
\]
Remember that neutrals are lambda-terms of the form $x\;a_1\;\dots\;a_n$ with
$x$ being a variable (the only one) declared in the stoup. The accumulator $g$
in the definition of $\netoF'$ is intended to collect the arguments
$a_i,\dots,a_n$ that have already been seen. So when a new argument
$a$ appears, which is a normal form, this is immediately
translated to an $\inv$-phase derivation via
$\nftoI$ and then pushed on top of the accumulator using the left rule
$\loL$. The accumulator $f$ in the definition of $\Ftone'$ serves a
similar purpose. The translations $\netoF$ and $\Ftone$ are then
easily definable:
\[\small
\begin{array}{l}
  \netoF : (\ne{A}{\Gamma}{C}) \to (\stseqF{A}{\Gamma}{C}) \\
  \netoF\;f = \netoF'\;f\;\ax
\end{array}
\qquad\qquad
\begin{array}{l}
  \Ftone : (\stseqF{A}{\Gamma}{C}) \to (\ne{A}{\Gamma}{C}) \\
  \Ftone\;f = \Ftone'\;\ax\; f
\end{array}
\]
These translations form an isomorphism. The crucial lemma for proving
this is: given $f : \ne{A}{\Gamma}{B}$ and $g :
\stseqF{B}{\Delta}{C}$, we have $\netoF\;(\Ftone'\;f\;g) =
\netoF'\;f\;g$ and $\Ftone\;(\netoF'\;f\;g) = \Ftone'\;f\;g$.

\section{Losing Skewness and How to Restore It}
\label{sec:complete}

The free skew prounital closed category $\FSkPCl(\At)$ on a set of atoms $\At$ is left normal,
which means that its skew aspect is superfluous. In other words,
$\FSkPCl(\At)$ is also the free (non-skew) prounital closed category
on $\At$. An advantage of our proof theoretic analysis is that left-normality can be proved in any one of the equivalent calculi of
Section~\ref{sec:free}. Left-normality is a simple observation in the
sequent calculus, while it is not clear how to derive it directly
in the categorical calculus of Section~\ref{sec:catcalc}.

First we notice that left-normality, defined as the invertibility of
the derivable map $\jY$ of (\ref{eq:ln}), is translated to the
invertibility of the passivation rule $\uf$ in the sequent calculus of
Section~\ref{sec:seqcalc}. In other words, $\jY$ is invertible up to
$\doteq$ in the categorical calculus if and only if $\uf$ is
invertible up to $\circeq$ in the sequent calculus. Then we show that
$\uf$ has as inverse the admissible rule $\ufinv$:
\begin{equation}\label{eq:ufinv}\small
\infer[\ufinv]{\stseq{A}{\Gamma}{C}}{
  \stseq{\n}{A,\Gamma}{C}
}
\end{equation}
This is defined by induction on the given derivation $f :
\stseq{\n}{A,\Gamma}{C}$. There are only two possible cases: if $f =
\uf\;f'$, define $\ufinv\;f = f'$; if $f = \loR\;f'$, define
$\ufinv\;f = \loR\;(\ufinv\;f')$.

An important consequence of left-normality is that all calculi
described in Sections~\ref{sec:seqcalc}--\ref{sec:norm} admit a
presentation without the stoup and the $\uf$ rule. In particular, the
natural deduction calculus of Section~\ref{sec:natded} is equivalent
to (non-skew) planar simply-typed lambda-calculus~\cite{Abr:temper,Zei:theltf}.
The categorical calculus of \ref{sec:catcalc} also admits a stoup-free
version where sequents take the form $\tto A$ where $A$ is a formula. The inference rules are 
\begin{equation}\label{eq:catcalc-stoupfree}\small
\renewcommand{\arraystretch}{2}
\begin{array}{c@{\quad \quad}c@{\quad \quad}c}
&
\infer[\dcomp']{\tto C}{\tto B & \tto B \lo C}
&
\\
\infer[j]{ \tto A \lo A}{}
&
\infer[i']{\tto (A \lo B) \lo B}{\tto A}
&
\infer[L']{\tto (B \lo C) \lo ((A \lo B) \lo (A \lo C))}{}
\end{array}
\end{equation}
Under the Curry-Howard correspondence, this is the combinatory logic
capturing planar lambda-calculus: $\dcomp'$ is
application, $j$ is the $I$-combinator, and $L'$ is the
$B$-combinator, while the operation $i'$ replaces the $C$-combinator of 
$BCI$ combinatory logic \cite{MP:notapp} and is needed in the absence of symmetry.

A natural question arises: why did we bother including the stoup in
our calculi in the first place? There are two reasons behind our
choice to include the stoup.

First, in the future we plan to extend the skew calculi described in
this paper with other connectives, such as unit and
tensor. We already know from previous work on the proof theory of skew
monoidal categories~\cite{UVZ:seqcsm} that the extended calculi will
not be left-normal, so we will not be able to discard the stoup. We
believe that a thorough investigation of the normalization procedures
of Sections~\ref{sec:focus} and \ref{sec:norm}, which work in the
presence of the stoup, is a stepping stone towards the development of
normalization functions for more involved calculi with additional
connectives.

Second, the left-normality of $\FSkPCl(\At)$ arises from the fact that
we are considering the free skew prounital closed category on a
\emph{set}.
In other words, it corresponds to a left
adjoint to the forgetful functor between the category of skew
prounital closed categories and strict prounital closed functors and
the category of sets and functions.
From a categorical point of view,
there is no good reason to privilege the category of sets and
functions in this picture. The next subsection is devoted to the
study of the free skew prounital closed category $\FskPCl(\M)$ on a
skew multicategory $\M$. The category $\FSkPCl(\At)$ arises as a
particular instance of the latter more general
construction. Crucially, $\FSkPCl(\M)$ is generally not left-normal.

\subsection{The Free Skew Prounital Closed Category on a Skew Multicategory}

We start by recollecting Bourke and Lack's notion of skew
multicategory~\cite{BL:multi}.  We slightly reformulate Bourke and
Lack's definition to make its relationship to the sequent calculus of
Section~\ref{sec:seqcalc} more direct.  Skew multicategories are
similar to the multicategories of Lambek \cite{Lam:ded2} (also
known as colored (non-symmetric) operads), but instead use an optional
object paired with a list of objects as the domain of a multimap,
rather than just a list of objects.

A \emph{skew multicategory} $\M$ consists of a set $\M_0$ of objects and, for
any optional object $S$, list of objects $\Gamma$ and object $C$ in $\M_0$, a
set $\stseqM{S}{\Gamma} C$ of multimaps whereby a multimap is called \emph{loose} if $S$ is empty and \emph{tight} if $S$ is an object. For any object $A$, there is
an identity multimap $\id \in \stseqM{A}{~} A$.
There are two composition operations $\mathsf{s}\circ : \stseqM{A}{\Delta}{C}
\times \stseqM{S}{\Gamma}{A} \to \stseqM{S}{\Gamma, \Delta}{C}$
and $\mathsf{c}\circ : \stseqM{S}{\Delta_0,A,\Delta_1}{C}
\times \stseqM{\n}{\Gamma}{A} \to \stseqM{S}{\Delta_0, \Gamma, \Delta_1}{C}$
and a loosening operation $\loosen : \stseqM{A}{\Gamma}{C} \to 
\stseqM{\n}{A, \Gamma}{C}$
satisfying a large number of equations, expressing unitality of
identity wrt.\ composition, associativity of composition, commutativity of
parallel cuts and commutativity of composition and loosening. See the
whole list of equations in our previous work~\cite{UVZ:seqcsm}.


A \emph{skew multifunctor} $G$ between skew multicategories $\M$ and
$\M'$ consists of a function $G_0$ sending objects of $\M$ to objects
of $\M'$ and a function
$G_1 : \stseqM{S}{\Gamma}{C} \to \stseqMpr{G_0 S}{G_0 \Gamma}{G_0 C}$
preserving identity, composition and loosening. Here $G_0$ is extended to optional objects and lists of objects by $G_0\, S = \n$ if
$S = \n$ and $G_0\,S = G_0\, A$ if $S = A$. Similarly,
$G_0(A_1, \ldots, A_n) = G_0\, A_1, \ldots, G_0\, A_n$. Skew
multicategories and skew multifunctors form a
category. There exists
a forgetful functor $U$ between the category of skew prounital
closed categories and the latter category.
Given a skew prounital
closed category $\C$, we define $U\;\C$ as the skew multicategory with
the same objects as $\C$ and with the multihomset $\stseqU{S}{\Gamma}{C}$
given by $\C(S, \asem{\Gamma}{C})$. From the structure of $\C$, using properties of the
interpretation $\asem{\Gamma}C$, one defines the identity, composition and loosening of $U\;\C$.


The \emph{free} skew prounital closed category on a skew multicategory
$\M$ is then a skew prounital closed category $\FskPCl(\M)$ equipped
with a skew multifunctor $\iota: \M \to U(\FSkPCl(\M))$. For any other
skew prounital closed category $\C$ with a skew multifunctor $G : \M \to
U\;\C$, there must exist a unique strict prounital closed functor $\bar{G}
: \FSkPCl(\M) \to \C$ compatible with $\iota$.

Let $\M$ be a skew multicategory. We construct a categorical calculus
presenting the free skew prounital closed category on $\M$. We then
proceed to describe an equivalent cut-free sequent calculus.

\paragraph{Categorical Calculus}

The formulae are given by objects $X \in \M_0$ (atomic formulae) and $A \lo B$
for any formulae $A$, $B$. The inference rules are the same as in
(\ref{eq:catcalc}), supplemented with an additional inference
rule
\[\small
\infer[\iota]{T \tto \asem{\Phi}{Z}}{
  \stseqM{T}{\Phi}{Z}
}
\]
where $T$ is an optional atom, $\Phi$ is a list of atoms and $Z$ is an atom. 
The equational theory $\doteq$ from (\ref{eq:doteq}) is extended with new generating
equations expressing the fact that $\iota$ is a skew multifunctor
between $\M$ and $U(\FSkPCl(\M))$.

\paragraph{Cut-Free Sequent Calculus}

The inference rules are those given in
(\ref{eq:seqcalc}) minus the rules $\ax$ and $\uf$ plus two new rules
\[\small
\infer[\iota]{\stseq{T}{\Phi}{Z}}{
  \stseqM{T}{\Phi}{Z}
}
\qquad\qquad
\infer[\loC]{\stseq{S}{\Delta_0,A \lo B,\Gamma,\Delta_1}{C}}{
  \stseq{\n}{\Gamma}{A}
  &
  \stseq{S}{\Delta_0,B,\Delta_1}{C}
}
\]
The rule $\loC$ was derivable using cut in the sequent calculus of
Section~\ref{sec:seqcalc}, as we showed in (\ref{eq:loC}).  Here it
is needed as a primitive rule to achieve cut admissibility. 
From the presence of a map 
$f \in \stseqM{X}{Y}{Z}$ in the base skew multicategory
$\M$, we need to be able to derive, e.g., the sequent $\stseq{X}{A \lo
  Y,A}{Z}$, which in the categorical calculus is derivable as follows:
\[\small
\infer[\dcomp]{X \tto (A \lo Y) \lo (A \lo Z)}{
  \infer[\iota]{X \tto Y \lo Z}{
    \stseqM[f]{X}{Y}{Z}
  }
  &
  \infer[L]{Y \lo Z \tto (A \lo Y) \lo (A \lo Z)}{}
}
\]

The equational theory on derivations is obtained from the congruence
$\circeq$ of (\ref{eq:circeq}) 
by adding the following generating equations:
\[\small
\begin{array}{c@{\;\,}l}
 \multicolumn{2}{l}{\textrm{(preservation of $\id$ and $\loosen$ by $\iota$)}} \\
 \ax_X \circeq \iota\;(\id_X) & \\
 \uf\;(\iota\;f) \circeq \iota\;(\loosen\;f) 
 & (\text{for } f \in \stseqM{X}{\Phi}{Z})
 \\[9pt]
  \multicolumn{2}{l}{\textrm{(commutative conversions of $\loC$)}} \\
  \loC\;(f,\loR\;g) \circeq \loR\;(\loC\;(f,g))
  & (\text{for } f : \stseq{\n}{\Gamma}{A'}, g : \stseq{S}{\Delta_0,B',\Delta_1,A}{B})
  \\
 \uf\;(\loC\;(f,g)) \circeq \loC\;(f,\uf\;g) 
 & (\text{for } f : \stseq{\n}{\Gamma}{A}, g : \stseq{A'}{\Delta_0,B,\Delta_1}{C})
 \\
 \uf\;(\loL\;(f,g)) \circeq \loC\;(f,\uf\;g) 
 & (\text{for } f : \stseq{\n}{\Gamma}{A}, g : \stseq{B}{\Delta}{C})
 \\
  \loC\;(f,\loL(g,h)) \circeq \loL\;(g,\loC\;(f,h))
  & (\text{for } f : \stseq{\n}{\Gamma}{A}, g : \stseq{\n}{\Gamma'}{A'}, h : \stseq{B'}{\Delta_0,B,\Delta_1}{C})
  \\
  \loC\;(f,\loL(g,h)) \circeq \loL\;(\loC\;(f,g),h)
  & (\text{for } f : \stseq{\n}{\Gamma}{A}, g : \stseq{\n}{\Delta_0,B,\Delta_1}{A'}, h : \stseq{B'}{\Delta}{C})
  \\
  \loC\;(f,\loC(g,h)) \circeq \loC\;(g,\loC\;(f,h)) 
  & (\text{for } f : \stseq{\n}{\Gamma}{A}, g : \stseq{\n}{\Gamma'}{A'}, h : \stseq{S}{\Delta_0,B,\Delta_1,B',\Delta_2}{C})
  \\
  \loC\;(f,\loC(g,h)) \circeq \loC\;(\loC\;(f,g),h) 
  & (\text{for } f : \stseq{\n}{\Gamma}{A}, g : \stseq{\n}{\Delta_0,B,\Delta_1}{A'}, h : \stseq{S}{\Delta_2,B',\Delta_3}{C})
\end{array}
\]

Thanks to the presence of the primitive rule $\loC$, the two cut rules
in (\ref{eq:cut}) are admissible in this sequent calculus. In this
case, they need to be defined by mutual induction with another cut
rule
\[\small
\infer[\ccutj]{\stseq S {\Delta_0, A',\Gamma, \Delta_1} C}{
  \stseq {A'} {\Gamma} A
  &
  \stseq S {\Delta_0, A, \Delta_1} C
}
\]
In the sequent calculus of Section~\ref{sec:seqcalc}, the rule
$\ccutj$ is definable by first applying $\uf$ to the first premise and
then using $\ccut$.  In the new sequent calculus of the current
section, we have to define it simultaneously with $\scut$ and $\ccut$
because of the added cases for the added primitive rules. It is
possible to prove that the embedding $\iota$ is a skew multifunctor,
in particular it preserves the cut operations.

Notice that the sequent calculus is generally not left-normal. In
fact, an attempt to prove the admissibility of the rule $\ufinv$ of
(\ref{eq:ufinv}) fails when the premise is of the form
$\iota\;f$ for some $f \in \stseqM{\n}{X,\Phi}{Z}$. E.g., we 
may well have a map in $\stseqM{\n}{X}{Z}$ for some $X$ and $Z$ without there being any map in $\stseqM{X}{}{Z}$. Therefore the stoup
cannot be discarded.

The categorical calculus and the sequent calculus are equivalent. It is
possible to construct functions $\sound$ and $\cmplt$ translating
between the two calculi, and show that they form an isomorphism
up to the extended equivalence relations $\doteq$ and
$\circeq$. 


\paragraph{Focused derivations} The focused subcalculus uses 
that the $\ax$ and $\pass$ rules of the sequent
calculus are admissible from $\id$ and $\loosen$ (crucially because
the presence of $\loC$ makes it possible to commute $\pass$ and $\loL$). 
It has the inference rules from (\ref{eq:foccalc}) minus the rules
$\pass$ and $\ax$ plus two new rules
\[\small
\infer[\iota]{\stseqF{T}{\Phi}{Z}}{
  \stseqM{T}{\Phi}{Z}
}
\qquad
\infer[\loC]{\stseqF{T}{\Psi, A \lo B,\Gamma,\Delta}{C}}{
  \stseqI{\n}{\Gamma}{A}
  &
  \stseqF{T}{\Psi,B,\Delta}{C}
}
\]
Notice that, in the rule $\loC$, $T$ is restricted to be an
optional atom and $\Psi$ a list of atoms.
Notice also that the passivation phase is trivial because we
have removed the rule $\pass$.


\subsection{Starting From a Skew Multigraph}

The free skew prounital closed category $\FskPCl(\M)$ over a multicategory $\M$ is
special in that we can have a cut-free sequent calculus where the use
of the generating multimaps is confined to ``direct import'' by
$\iota$. In fact, no structural rules (neither any cut rules nor
$\pass$ or $\ax$) are needed beyond the degree that they are readily
available to us in the form of composition, loosening and identity in the base
skew multicategory $\M$ where they also satisfy the skew multicategory
equations. This is possible because the cut rules happen to be
admissible from the sound rule $\loC$ that we may choose to take as primitive.

This approach is not robust for extensions with further connectives; we cannot
have a similar cut-free sequent calculus for the free skew (unital) closed
category $\mathbf{FskCl}(\M)$: from $\stseqM{\n}{~}{Y}$ and $\stseqM{X}{Y}{Z}$, we must be
able to derive $\stseq{X}{\I}{Z}$, but for this we need $\ccut$ as a
primitive rule (together with $\pass$) since, differently from $\lo$,
it is unsound to introduce $\I$ into the passive context. However, as
soon as we introduce primitive cut rules (it suffices to take $\scut$
and $\ccut$ as primitive), we also need to introduce (i) equations
stating that $\iota$ preserves compositions as cuts and (ii) also the
skew multicategory equations for $\scut$, $\ccut$, $\ax$ and
$\pass$. The equations (i) can be dispensed with if we start with a
\emph{skew multigraph} $(\Var,\G)$ (of \emph{atoms} and \emph{definite clauses}) rather than a skew multicategory $\M$, so
that composition as well as the identities and loosening are only
available in terms of $\scut$, $\ccut$, $\ax$ and $\pass$. We
conjecture that the equations (ii) can then also be avoided in a
focused subcalculus with all the inference rules from (\ref{eq:foccalc}) plus
the rule
\[
\small
\infer[\iota']{\stseqF{T}{\Gamma_1,\ldots,\Gamma_n,\Delta}{C}}{
 \stseqI{\n}{\Gamma_1}{Y_1}
 & 
 \ldots
 &
 \stseqI{\n}{\Gamma_n}{Y_n}
 &
 \stseqG{T}{Y_1,\ldots,Y_n}{X}
 &
 \stseqF{X}{\Delta}{C}
}
\]
which packages a particular combination of  $\iota$ and $\scut$ and
$\ccut$ inferences.

\section{Conclusions and Future Work}
\label{sec:concl}

We presented several equivalent presentations of the free skew
prounital closed category on a set $\At$. We showed that these
correspond to a skew variant of the planar fragment of linear typed
lambda-calculus. We constructed two calculi of normal forms: a focused
sequent calculus and a normal natural deduction calculus.  These solve
the coherence problem for skew prounital closed categories by fully
characterizing the homsets of $\FSkPCl(\At)$. The latter category is
left-normal, meaning that its skew aspect is redundant. We restored
the skewness by studying deductive systems for the free skew prounital
closed category on a skew multicategory and showing that the
latter is generally not left-normal.

The development presented in the paper has been fully formalized in
the dependently typed programming language Agda. Our Agda
formalization also includes a similar proof theoretic analysis of the
free \emph{skew closed category} on a set, in which the element set
functor $J$ is replaced by a unit object $\I$~\cite{Str:skecc}. The
primitive rules of the cut-free sequent calculus of skew closed
categories also include left and right introduction rule for the unit
$\I$, where again the left rule acts only on the unit in the
stoup. Similarly, the natural deduction calculus has introduction and
elimination rules for $\I$. Our reduction-free normalization
procedures can be adapted to the skew closed case without much
difficulty.

In the future, we plan to extend the work of this paper and our
previous work on the sequent calculus of the Tamari
order~\cite{Zei:seqcsa} and of skew monoidal
categories~\cite{UVZ:seqcsm,UVZ:protpn} to a proof theoretic
investigation of skew monoidal closed categories, i.e. including unit
$\I$, tensor $\ot$ and internal hom $\lo$ related by an adjunction
${-} \ot B \dashv B \lo {-}$. We already know from our previous work
that the corresponding sequent calculus would not be left-normal. We
conjecture that the free skew monoidal closed category on $\At$
corresponds to a skew variant of the $(\I,\ot,\lo)$ fragment of
noncommutative intuitionistic linear logic~\cite{Abr:noncil}. It is
currently not clear how to extend the normalization procedures of this
paper to the skew monoidal closed case, in particular normalization by
evaluation, which has not been studied in the planar (or even linear)
fragment of lambda-calculus. Inspiration could come from the
normalization by hereditary substitution algorithm of Watkins et
al.~\cite{WCPW:conlfp} for the propositional fragment of their concurrent logical
framework.


\paragraph{Acknowledgments.}

We thank the anonymous referees for extremely valuable comments.
T.U.\ was supported by the Icelandic Research Fund grant
no.~196323-052 and the Estonian Ministry of Education and Research
institutional research grant no.~IUT33-13. N.V.\ was supported by the
ESF funded Estonian IT Academy research measure (project
2014-2020.4.05.19-0001).





\newcommand{\doi}[1]{\href{https://doi.org/#1}{doi: #1}}

\end{document}